\documentclass[submission, Phys]{SciPost}

\hypersetup{
    colorlinks,
    linkcolor={red!50!black},
    citecolor={blue!50!black},
    urlcolor={blue!80!black}
}

\usepackage[bitstream-charter]{mathdesign}
\DeclareSymbolFont{usualmathcal}{OMS}{cmsy}{m}{n}
\DeclareSymbolFontAlphabet{\mathcal}{usualmathcal}

\begin{document}

\begin{center}{\Large \textbf{Transparency-engineered SQUID cells for Kerr-free three-wave-mixing Josephson metamaterials}}\end{center}

\begin{center}
Claudio Guarcello\textsuperscript{1,2,3$\star$}, A. Mert Bozkurt\textsuperscript{4}, Carlo Barone\textsuperscript{1,2,3}, Giovanni Filatrella\textsuperscript{5}, Alessandro Bruno\textsuperscript{4}, and Sergio Pagano\textsuperscript{1,2,3}
\end{center}

\begin{center}
{\bf 1} Dipartimento di Fisica ``E.~R.~Caianiello'', Universit\`{a} degli Studi di Salerno, I-84084 Fisciano, Salerno, Italy
\\
{\bf 2} INFN, Sezione di Napoli, Gruppo Collegato di Salerno - Complesso Universitario di Monte S. Angelo, I-80126 Napoli, Italy
\\
{\bf 3} Consiglio Nazionale delle Ricerche, CNR-SPIN, c/o Università di Salerno- via Giovanni Paolo II, 132, Fisciano (SA) 84084, Italy
\\
{\bf 4} QuantWare, Molengraaffsingel 8, 2629 JD Delft, The Netherlands
\\
{\bf 5} Department of Sciences and Technologies, University of Sannio, via de Sanctis, Benevento, I-82100, Italy

${}^\star$ {\small \sf cguarcello@unisa.it}
\end{center}

\begin{center}
\today
\end{center}


\section*{Abstract}
{\bf
We introduce a transparency-engineered rf-SQUID cell for Kerr-free three-wave-mixing Josephson metamaterials. This design replaces the conventional tunnel-junction element with an effective Josephson element formed by two junctions in series, yielding a non-sinusoidal energy-phase relation that can be used as a tunable nonlinear design resource. We show that the asymmetry between the two series junctions and the applied flux bias provide independent control over the local expansion of the rf-SQUID potential, enabling operating points where the leading quartic Kerr term is suppressed while the cubic nonlinearity remains finite. We derive the corresponding Kerr-free condition, identify the resulting operating ridge in parameter space, and analyze the local stability and passive-matching constraints that bound its physically accessible portion. Our results provide a compact unit-cell design principle for three-wave-mixing Josephson metamaterials and suggest a route toward Kerr-suppressed Josephson traveling-wave parametric amplifiers.
}

\vspace{10pt}
\noindent\rule{\textwidth}{1pt}
\tableofcontents\thispagestyle{fancy}
\noindent\rule{\textwidth}{1pt}
\vspace{10pt}

\section{Introduction}

Superconducting Josephson circuits provide a versatile platform for quantum-limited microwave amplification and quantum measurement~\cite{Clerk2010,Roy2016,Aumentado2020}. When arranged as nonlinear transmission-line metamaterials, Josephson junction (JJ) arrays enable engineered wave propagation, parametric wave mixing, and broadband amplification~\cite{Fasolo2019,Esposito2021}. A prominent realization is the Josephson traveling-wave parametric amplifier (JTWPA), where an extended nonlinear transmission line provides broadband low-noise microwave gain for multiplexed qubit readout and quantum measurement architectures. Compared with resonant Josephson parametric amplifiers, traveling-wave implementations can offer multi-GHz bandwidth, higher saturation power, and compatibility with large-scale frequency multiplexing, making them especially attractive for scalable superconducting quantum technologies~\cite{Bell2015,Zorin2016,Malnou2021,Ranadive2022,Gaydamachenko2025,Bell2026,Wang2025}. These advantages, however, come at the cost of a delicate interplay between nonlinearity, dispersion, and phase matching along an extended distributed medium~\cite{OPeatain2023,Bell2026}.

Standard Josephson implementations rely on either Kerr-type four-wave mixing (4WM) or flux-biased three-wave mixing (3WM) in composite nonlinear cells. In conventional 4WM JTWPAs, the same Kerr-type nonlinearity that enables amplification also induces self- and cross-phase modulation, pump-dependent phase shifts, and gain distortions, thereby making phase matching more challenging and often requiring elaborate dispersion-engineering strategies~\cite{Bell2015,OPeatain2023,Bell2026}. A central design problem is therefore how to obtain a strong and controllable cubic nonlinearity while suppressing the unwanted quartic Kerr contribution. This issue has become increasingly central in the broader development of superconducting parametric devices, where the goal is no longer simply to enhance nonlinearity, but to engineer it selectively.

A natural route to 3WM in Josephson traveling-wave devices is provided by flux-biased \textit{rf}-SQUID arrays.
\footnote{Here and in the following, \textit{rf}-SQUID is used in the sense common in the JTWPA literature, namely as a shorthand for a flux-biased single-junction superconducting loop used as the elementary nonlinear cell of a 3WM Josephson metamaterial, independently of the specific galvanic or non-galvanic bias/readout implementation.}
In these systems, the external flux tunes the operating point of the elementary cell within its nonlinear potential landscape, thereby controlling the balance between lower- and higher-order nonlinearities and enabling regimes favorable to 3WM amplification.~\cite{Zorin2016,Zorin2017,Miano2019,Zorin2021,Greco2021}. 
Over the last few years, this line of work has established rf-SQUID metamaterials as a concrete platform for high-dynamic-range Josephson amplification, while also clarifying the importance of fabrication spread, dispersion engineering, and phase-matching strategies in long arrays~\cite{Kissling2023,Giachero2022,Gaydamachenko2025}. In parallel, the SNAIL paradigm and related Kerr-free operation schemes have shown that suitably engineered Josephson potentials can retain a useful cubic nonlinearity while minimizing or suppressing the residual quartic Kerr term, leading to cleaner pumped operations and improved dynamic range~\cite{Frattini2017,Sivak2019,Ranadive2022,Roudsari2023}.

More recently, similar ideas have begun to appear also in single-junction settings based on engineered or skewed current--phase relations (CPRs), where the local Josephson energy landscape is tailored so as to produce third-order sweet spots without relying on more elaborate multi-junction loop architectures~\cite{Bozkurt2023,RezaFrolov2026}. These developments make clear that non-sinusoidal Josephson energy--phase relations (EPRs) should not be viewed merely as perturbations of otherwise standard devices, but rather as a real design resource. Recent works have explored how non-sinusoidal CPRs modify the behavior of JTWPAs, with particular emphasis on gain, stability, and the onset of complex dynamical regimes~\cite{Guarcello2024,Guarcello2025,Guarcello2026}. However, they addressed primarily the distributed dynamics of a given nonlinear medium. The focus of the present work is instead on the design of the nonlinear response of a single rf-SQUID cell, with the aim of obtaining an explicit Kerr-free 3WM operating manifold already at the level of the elementary building block.

Here we develop a route based on transparency engineering of a single-junction rf-SQUID cell. The resulting circuit element, which we call the TRAIL for TRansparency-engineered and Asymmetry-tuned Inductive Loop, provides a compact flux-biased building block for Kerr-free 3WM Josephson metamaterials.
The physical idea is close in spirit to recent approaches that engineer non-sinusoidal EPRs, but here the target is finding a relevant operating \emph{ridge} in parameter space along which the quartic coefficient vanishes while the cubic coefficient remains appreciable. The core idea is to replace the usual tunnel-junction with an effective high-transparency Josephson element obtained from two junctions in series, following the energy--phase engineering strategy of Ref.~\cite{Bozkurt2023}. This series-junction approach has already proved versatile well beyond its original formulation. Closely related implementations have been used for voltage-controlled harmonic synthesis in hybrid Josephson circuits, for hybrid rhombi with tunable \(\cos(2\phi)\) and diode regimes, for tunable double-junction transmons with resolvable higher harmonics, and for Fourier-engineered \(\cos(2\phi)\) protected qubits, showing that effective-transparency engineering and its multijunction generalizations provide a broadly applicable route to tailored Josephson EPRs~\cite{Banszerus2024PRL,Banszerus2025PRX,Shagalov2025,Zhurbina2026}. When embedded in an \textit{rf}-SQUID loop, the resulting effective Josephson element provides a controllable non-sinusoidal CPR, and the local expansion of the corresponding total potential can then be tuned through the asymmetry parameter and the applied magnetic flux. Compared with multi-junction asymmetric loop elements, the proposed architecture keeps the unit cell compact and shifts a substantial part of the nonlinearity engineering to the effective transparency parameter.

This point also clarifies the novelty of the present work. Here, the focus shifts from distributed-amplifier behavior to the nonlinear design of the elementary cell. Moreover, in contrast with recent Kerr-free proposals based on $\phi_0$-junction physics~\cite{RezaFrolov2026}, the mechanism discussed here relies on an effective transparency-engineered Josephson element embedded in an \textit{rf}-SQUID loop. The resulting framework yields a Kerr-free design ridge, the associated local stability condition, and a passive-matching threshold relevant for JTWPA implementation.

The manuscript is organized as follows. In Sec.~\ref{sec:effective_element} we introduce the effective transparent Josephson element and derive the \textit{rf}-SQUID potential. In Sec.~\ref{sec:local} we compute the local coefficients $c_2$, $c_3$, and $c_4$ around the equilibrium point and define the Kerr-free design strategy. In Sec.~\ref{sec:kfr} we translate the single-cell result into a JTWPA design principle. In Sec.~\ref{sec:matching} we discuss passive matching of the unit cell and the emergence of a critical transparency. We conclude in Sec.~\ref{sec:conclusions}.

\begin{figure}[t]
  \centering
  \includegraphics[width=\columnwidth]{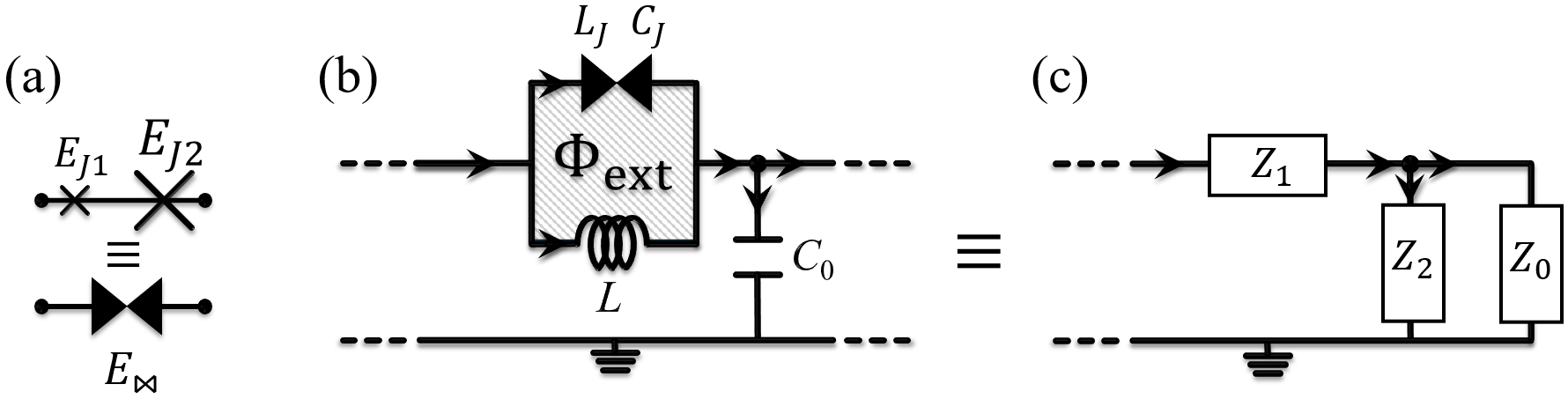}
\caption{Schematic construction of the TRAIL cell and its equivalent \(\pi\)-section model. (a) Two Josephson junctions in series, with energies \(E_{J_1}\) and \(E_{J_2}\), define an effective transparent Josephson element with energy--phase relation \(E_{\bowtie}\). (b) The effective element is embedded in a flux-biased inductive loop of inductance \(L\), with ground capacitance \(C_0\), yielding the elementary TRAIL cell; the linearized Josephson branch is described by the differential inductance \(L_J\) and capacitance \(C_J\). (c) Equivalent \(\pi\)-cell representation used for passive matching, with series impedance \(Z_1\), ground impedance \(Z_2\), and line impedance \(Z_0\).}
  \label{fig:device}
\end{figure}

\subsection{Effective transparent Josephson element}
\label{sec:effective_element}

We consider two JJs in series, see Fig.~\ref{fig:device}(a), with
Josephson energies
\begin{equation}
E_{J_1}=\frac{\Phi_0 I_{c_1}}{2\pi},
\qquad
E_{J_2}=\frac{\Phi_0 I_{c_2}}{2\pi},
\end{equation}
with \(I_{c_i}\) denoting the critical current of the \(i\)-th junction
and \(\Phi_0\) the flux quantum. The intermediate superconducting island
between the two junctions introduces an additional internal phase degree
of freedom. In the classical regime, where both Josephson energies
dominate over the charging energy of the island,
\(E_{J_i}\gg E_C\), charging-induced quantum fluctuations of this
degree of freedom are suppressed. Provided that the internal mode
remains adiabatically slaved to the total phase, this variable can be
eliminated and the two-junction structure can be described as an
effective short Josephson element with CPR and EPR given, respectively,
by~\cite{Bozkurt2023}
\begin{align}
I_{\blacktriangleright\!\blacktriangleleft}(\varphi)
&=
\frac{2\pi}{\Phi_0}\frac{E_J\,\tau}{4}\,
\frac{\sin\varphi}{\sqrt{1-\tau\,\sin^2(\varphi/2)}}
\qquad\text{and}\qquad
E_{\blacktriangleright\!\blacktriangleleft}(\varphi)
=
-\,E_J\,\sqrt{1-\tau\,\sin^2\!\Bigl(\tfrac{\varphi}{2}\Bigr)},
\label{eq:effJJ}
\end{align}
where \(E_J=E_{J_1}+E_{J_2}\) and the effective transparency \(\tau\)
is controlled by the asymmetry of the two junctions. Indeed, introducing
the asymmetry parameter
\begin{equation}
\label{eq:alpha}
\alpha=\frac{I_{c_2}}{I_{c_1}}
      =\frac{E_{J_2}}{E_{J_1}},
\end{equation}
we obtain
\begin{equation}
\tau=\frac{4\alpha}{(1+\alpha)^2},
\label{eq:alpha_tau}
\end{equation}
i.e., the series-junction element interpolates between weakly and
strongly non-sinusoidal regimes as \(\alpha\) changes. 
In this work we restrict to $\alpha < 1$, for the case $\alpha > 1$ is obtained interchanging ``$1$'' and ``$2$''.
In particular,
\(\alpha\) provides a convenient design parameter that controls the
harmonic content of the effective CPR through the corresponding EPR in
Eq.~\eqref{eq:effJJ}.

The reduction to the effective CPR in
Eq.~\eqref{eq:effJJ} relies on a separation of dynamical scales that allows neglect of the currents through the resistor and capacitance, while the condition \(E_{J_i}\gg E_C\) suppresses charging-induced quantum fluctuations of the intermediate-island phase.
Under these hypothesis it is possible to eliminate the phase of the intermediate island.
In fact, the effective transparent Josephson element of Fig.  \ref{fig:device}(a) is ruled by an effective single phase dynamics $\varphi$ if the superconducting Josephson channel carries most of the current. 
This requires that the larger spectral components of the solution are sufficiently slow compared to the plasma and relaxation scales of each of the
two-junction subsystem. 
A simple and conservative estimate to only retain the Josephson current in each elementary junction is: 
\begin{equation}
\frac{V}{R} \ll I_c,    
\label{eq:Rcurrent}
\end{equation}
that is, a shunt current through the resistor much lower than the current through the Josephson element. 
Using the Josephson relation and a sinusoidal waveform and phase excursion of order unity such that $d\varphi /dt \simeq \omega$, Eq.  (\ref{eq:Rcurrent}) gives:
\begin{equation}
V= \frac{\hbar}{2e}\frac{d \varphi}{dt} \simeq \frac{\hbar}{2e } \omega \ll R I_c 
\end{equation}
and therefore the current through the resistor can be neglected respect to the Josephson supercurrent if the oscillations are slower than the characteristic frequency $\omega_c$:
\begin{equation}
\omega \ll \frac{2eR I_c}{\hbar} \equiv \omega_c.
\label{eq:omega_c_condtion}
\end{equation}
An analogous condition follows from the current through the capacitive branch:
\begin{equation}
    C \frac{dV}{dt} = C \frac{\hbar}{2e} \frac{d^2\varphi}{dt^2} \ll I_c
    \label{eq:Ccurrent}
\end{equation}
that, for the same approximation of a unitary sinusoidal waveform ($d^2\varphi /dt^2 \simeq \omega^2$), reads 
\begin{equation}
    \omega^2 \ll \frac{2e}{\hbar}\frac{I_c}{C} \equiv \omega_p^2
\label{eq:omega_p_condition}
\end{equation}
(the oscillations of the phase are much slower than the plasma oscillations $I_c  2e/C \hbar$ of the junction.)
These conditions should be understood as simple estimates rather than as a complete normal-mode analysis of the circuit.

For an array of junctions, the conditions (\ref{eq:omega_c_condtion},\ref{eq:omega_p_condition}) should be checked for every JJ, with the lowest critical current setting the most restrictive bound. 
If the junctions are fabricated on the same chip, using the same fabrication process, with the same specific critical-current density, capacitance, and resistance per unit area, both $\omega_c$ and $\omega_p$ are area independent.

A word is in order concerning the connection between the present device
and more traditional series-connected JJs. Historically, JJ arrays have
been studied primarily for their high-frequency properties, often in
regimes where the operating frequencies approach the characteristic or
plasma frequencies. Under such conditions,
Eqs.~(\ref{eq:omega_c_condtion}) and
(\ref{eq:omega_p_condition}) are inherently not satisfied. The
resistive and capacitive branches must therefore be retained, or at
least the resistive contribution must be included, as in studies of
multijunction overdamped series arrays and
SQUIDs~\cite{Hadley1988,Nappi1995,Wiesenfeld1996}. In this regime, the
internal phase becomes an independent dynamical degree of freedom, and
the composite element may develop a frequency-dependent response that
is not fully captured by the EPR and CPR in
Eq.~(\ref{eq:effJJ}).
More recently, microwave measurements on a related hybrid Josephson
rhombus have provided experimental support for treating, under
appropriate operating conditions, such a multijunction circuit as a
single effective Josephson element~\cite{Banszerus2025PRX}. A lumped
RCSJ model based on the ground-state CPR successfully reproduced
Shapiro-step measurements at \(4\,\mathrm{GHz}\), with agreement
reported over the experimentally explored range
\(1\!-\!8\,\mathrm{GHz}\). Since these measurements probe a strongly
driven, finite-voltage regime, they provide a nontrivial test of the
assumption that the internal degrees of freedom follow the collective
phase adiabatically.

In TRAIL operation, the signal is by definition weak, whereas the pump signal may induce significant phase oscillations, large and fast enough for the resistive and capacitive branches to become non negligible respect to the superconducting branches.
The reduced description therefore requires verification that the resistive and capacitive current components remain small compared with the Josephson currents $I_{c_1}$ and $I_{c_2}$, as suggested by the experiments in \cite{Banszerus2025PRX}.
If this is the case, the pure Josephson potential and its local expansion coefficients dominates. 
However, outside the adiabatic regime (\ref{eq:Rcurrent},\ref{eq:Ccurrent}), they may no longer provide a complete description of the finite-frequency impedance and amplifier response for the dynamical corrections due to the resistive and capacitive terms. 
A complete treatment of the full RCSJ dynamics is left for future work.

\subsection{rf-SQUID potential and operating points}

We now embed the effective Josephson element in a single-junction superconducting loop of non-negligible inductance \(L\),  see Fig.~\ref{fig:device}(b). Flux quantization relates the junction phase \(\varphi\) to the total magnetic flux \(\Phi\) threading the loop as~\cite{Clarke2004}
\begin{equation}
\varphi=2\pi\frac{\Phi}{\Phi_0}.
\label{eq:flux_quantization}
\end{equation}
The total flux is the sum of the externally applied flux and the self-induced contribution due to the circulating current, $\Phi=\Phi_{\rm ext}+L I_{\rm circ}$, so that 
\begin{equation}
I_{\rm circ}=\frac{\Phi-\Phi_{\rm ext}}{L}=\frac{\Phi_0}{2\pi L}\,(\varphi-2\pi\phi).
\label{eq:Icirc}
\end{equation}
As usual, the loop therefore contributes an inductive energy \(L I_{\rm circ}^2/2\), while the effective junction contributes the Josephson energy in Eq.~\eqref{eq:effJJ}.

It is convenient to normalize the total potential to \(E_{J_1}\) and to introduce the \emph{screening parameter}~\cite{Clarke2004}
\begin{equation}
\beta=\frac{2\pi }{\Phi_0}L I_{c_1}.
\label{eq:beta}
\end{equation}
For the representative values \(I_{c_1}=2~\mu\mathrm{A}\) and \(L_G=120~\mathrm{pH}\)~\cite{Pagano2022}, Eq.~\eqref{eq:beta} gives \(\beta\simeq 0.73\), which sets the relative weight of the inductive confinement with respect to the Josephson nonlinearity in the illustrative examples discussed below. The screening parameter \(\beta\) quantifies how strongly the circulating current generated in the loop feeds back on the total flux, and therefore controls the degree to which the inductive part of the SQUID dynamics competes with the Josephson nonlinearity~\cite{Clarke2004}. For \(\beta< 1\), the flux response is essentially single-valued and weakly screened, whereas for \(\beta\gtrsim 1\) the single-junction \textit{rf}-SQUID enters the familiar multistable regime in which hysteretic behavior can occur~\cite{Rifkin1976,Clarke2004,Bo2004,Guarcello2017,Guarcello2020}.

The dimensionless potential of the \textit{rf}-SQUID reads
\begin{equation}
\frac{U(\varphi)}{E_{J_1}}
=
\frac{(\varphi-2\pi\phi)^2}{2\beta}
-
(1+\alpha)\sqrt{1-\tau\,\sin^2\!\Bigl(\frac{\varphi}{2}\Bigr)}.
\label{eq:Usquid}
\end{equation}
Using the definition in Eq.~\eqref{eq:alpha_tau}, the Josephson contribution can be rewritten in the compact form
\begin{equation}
(1+\alpha)\sqrt{1-\tau\,\sin^2\!\Bigl(\frac{\varphi}{2}\Bigr)}
=
\sqrt{1+\alpha^2+2\alpha\cos\varphi}
\equiv
\sqrt{D(\varphi)}.
\label{eq:Didentity}
\end{equation}
The normalized \textit{rf}-SQUID potential therefore becomes
\begin{equation}
\frac{U(\varphi)}{E_{J_1}}
=
\frac{(\varphi-2\pi\phi)^2}{2\beta}
-
\sqrt{D(\varphi)}.
\label{eq:Usquid_compact}
\end{equation}

\begin{figure}[t]
  \centering
  \includegraphics[width=\columnwidth]{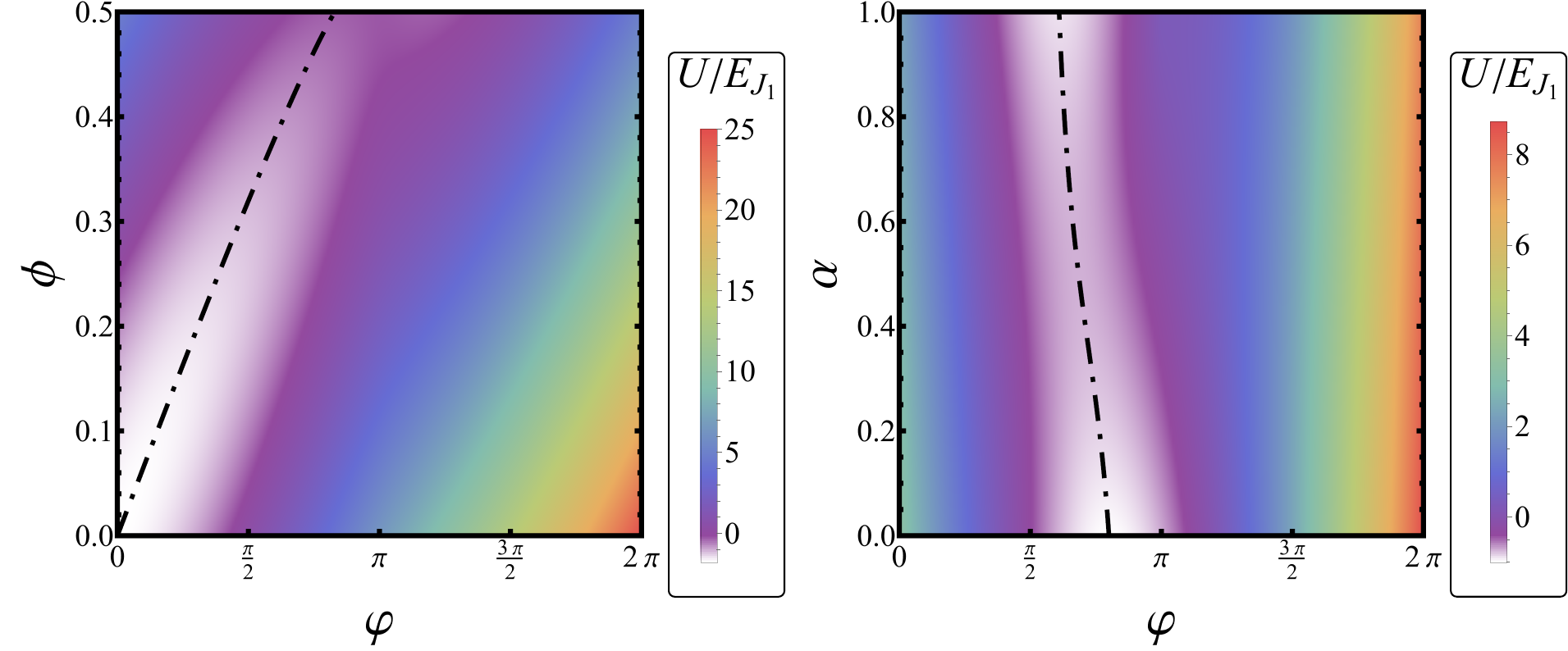}
  \caption{Normalized potential $U/E_{J_1}$ for the transparent junction built from two series JJs  for $\beta=0.73$. Panel (a): $U(\varphi,\phi)$ and $\alpha=0.75$; the black dot-dashed line marks $\varphi_{\min}(\phi)$. Panel (b): $U(\varphi,\alpha)$ and $\phi=0.4$; the black dot-dashed line marks $\varphi_{\min}(\alpha)$.}
  \label{fig:U-maps}
\end{figure}

In the weak-transparency limit \(\tau\ll 1\) (that is $E_{J_2} \ll E_{J_1}$), the Josephson contribution reduces to the conventional form up to an irrelevant additive constant. 
Indeed,
\begin{equation}
-(1+\alpha)
\sqrt{1-\tau\sin^2\left(\frac{\varphi}{2}\right)}
\simeq
-(1+\alpha)
\left[
1-\frac{\tau}{2}\sin^2\left(\frac{\varphi}{2}\right)
\right]
\simeq
-\frac{(1+\alpha)\tau}{4}\cos\varphi+\mathrm{const.}
\end{equation}
Using \(\tau=4\alpha/(1+\alpha)^2\), the normalized potential therefore reduces to
\begin{equation}
\frac{U(\varphi)}{E_{J1}}
\simeq
\frac{(\varphi-2\pi\phi)^2}{2\beta}
-
\frac{\alpha}{1+\alpha}\cos\varphi
+\text{const.}
\end{equation}
Thus, in this limit, the series element behaves as an effective cosinusoidal Josephson element with effective Josephson energy (and hence effective critical current)
\begin{equation}
E_J^{\mathrm{eff}}
=
\frac{E_J\tau}{4}
=
\frac{E_{J1}E_{J2}}{E_{J1}+E_{J2}}.
\end{equation}
Equation~\eqref{eq:Usquid} is the core of the model: it shows that the local nonlinear landscape of the unit cell is jointly controlled by the loop screening parameter \(\beta\), the junction asymmetry \(\alpha\), and the reduced external flux \(\phi\). In particular, \(\alpha\) governs the degree of non-sinusoidal behavior of the effective Josephson element, while \(\phi\) selects the operating point within the resulting potential landscape.

The stationary operating points are obtained from the condition \(\partial_\varphi U=0\), which yields
\begin{equation}
\frac{\varphi-2\pi\phi}{\beta}
+
\frac{\alpha\sin\varphi}{\sqrt{D(\varphi)}}=0.
\label{eq:stat}
\end{equation}
Solutions of Eq.~\eqref{eq:stat} define candidate bias points \(\varphi_{\min}(\beta,\alpha,\phi)\), and the physically relevant operating points are the locally stable minima of the potential, i.e., those for which \(\partial_\varphi^2 U>0\). In the next section we expand the potential around such minima and derive the local coefficients that determine the quadratic, cubic, and quartic nonlinear response of the unit cell.

The dependence of the normalized potential landscape, $U(\varphi)$, on the external flux and on the asymmetry parameter is illustrated in Fig.~\ref{fig:U-maps}. Panel~(a) shows how, at fixed \(\beta\) and \(\alpha\), varying \(\phi\) shifts the minimum of the \textit{rf}-SQUID potential along a well-defined trajectory \(\varphi_{\min}(\phi)\), traced with a dot-dashed white line. Panel~(b) shows that, at fixed \(\beta\) and \(\phi\), varying \(\alpha\) modifies not only the position of the minimum but also the overall shape and skewness of the local potential well. This already anticipates the central point of the present work: the combination of flux bias and series-junction asymmetry provides direct control over the local nonlinear coefficients of the elementary \textit{rf}-SQUID cell.

\section{Local expansion around the operating point}
\label{sec:local}

Let \(\varphi_{\min}\equiv\varphi_{\min}(\beta,\alpha,\phi)\) be a local minimum of the normalized potential \(u(\varphi)\equiv U(\varphi)/E_{J1}\) in Eq.~\eqref{eq:Usquid_compact}, namely a solution of the stationary condition
\begin{equation}
\frac{\varphi_{\min}-2\pi\phi}{\beta}
+
\frac{\alpha\sin\varphi_{\min}}{\sqrt{D_{\min}}}
=0,
\qquad\text{with}\qquad
D_{\min}\equiv 1+\alpha^2+2\alpha\cos\varphi_{\min}.
\label{eq:stat_min}
\end{equation}
Introducing the local coordinate \(\delta\varphi=\varphi-\varphi_{\min}\), we expand the normalized potential as
\begin{equation}
u(\varphi)
=
u(\varphi_{\min})
+
\sum_{k\geq 1}\frac{c_k}{k!}\,(\delta\varphi)^k,
\qquad
c_k=
\left.
\frac{d^k u}{d\varphi^k}
\right|_{\varphi=\varphi_{\min}}.
\label{eq:local_expansion}
\end{equation}
The coefficients \(c_k\) provide a compact local description of the effective nonlinearity experienced by the cell at the chosen bias point.
For the first few orders, one finds
\begin{align}
c_1 &=
\frac{\varphi_{\min}-2\pi\phi}{\beta}
+
\frac{\alpha\sin\varphi_{\min}}{\sqrt{D_{\min}}}
=0,
\label{eq:c1}
\\[4pt]
c_2 &=
\frac{1}{\beta}
+
\frac{\alpha}{2}\frac{\sqrt{D_{\min}}
}{D_{\min}^2}\Bigl[
2(1+\alpha^2)\cos\varphi_{\min}
+\alpha\bigl(3+\cos(2\varphi_{\min})\bigr)
\Bigr],
\label{eq:c2}
\\[4pt]
c_3 &=
-\frac{\alpha}{2}\,
\frac{\sqrt{D_{\min}}
}{D_{\min}^3}\Bigl[
2-\alpha^2+2\alpha^4
+2(\alpha+\alpha^3)\cos\varphi_{\min}
+\alpha^2\cos(2\varphi_{\min})
\Bigr]
\sin\varphi_{\min},
\label{eq:c3}
\\[4pt]
c_4 &=
-\frac{\alpha}{8}\,
\frac{\sqrt{D_{\min}}}{D_{\min}^4}
\Bigl\{
4(2+5\alpha^2+5\alpha^4+2\alpha^6)\cos\varphi_{\min}
+\alpha\Bigl[
28-21\alpha^2+28\alpha^4
\notag\\
&\hspace{2.2cm}
-4\bigl(1-9\alpha^2+\alpha^4\bigr)\cos(2\varphi_{\min})
+4(\alpha+\alpha^3)\cos(3\varphi_{\min})
+\alpha^2\cos(4\varphi_{\min})
\Bigr]
\Bigr\}.
\label{eq:c4}
\end{align}

\begin{figure}[t]
  \centering
  \includegraphics[width=\columnwidth]{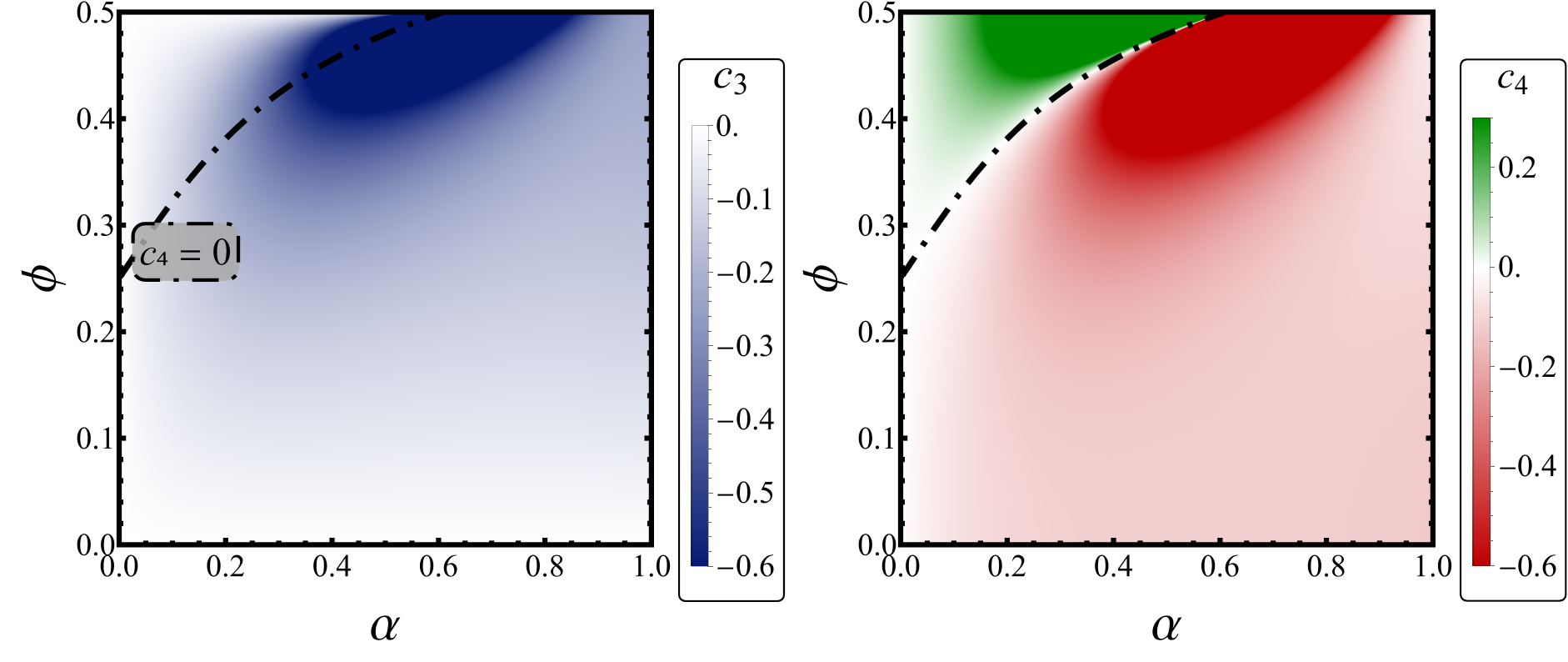}
  \caption{Cubic and quartic coefficients in the local expansion of the normalized potential
\(u\) for $\beta=0.73$. Panel (a): map of $c_3(\alpha,\phi)$. Panel (b): map of $c_4(\alpha,\phi)$. The black dot-dashed curve marks the Kerr-free ridge $c_4=0$.}
  \label{fig:c3c4-maps}
\end{figure}

The coefficients in Eq.~\eqref{eq:local_expansion} admit a direct physical interpretation. In the language of local potential expansions used for Kerr-free 3WM elements, the term proportional to \(c_3\) is the leading cubic contribution responsible for 3WM, whereas the term proportional to \(c_4\) is the leading quartic Kerr-type contribution~\cite{Frattini2017,RezaFrolov2026}. In the broader JTWPA literature, the same distinction is often phrased in terms of nonlinear coefficients controlling 3WM efficiency and Kerr-induced phase mismatch~\cite{Zorin2017,Miano2019,Ranadive2022}. 
Since the coefficients \(c_k\) are defined from the normalized potential \(u=U/E_{J1}\), one has
$U''(\varphi_{\min})=E_{J_1}c_2$.
The quadratic coefficient \(c_2\) therefore sets the local curvature of the well and the corresponding small-signal stiffness of the biased cell.
This can be seen directly from the local relation between current and phase,
\begin{equation}
I=\frac{2\pi}{\Phi_0}\,\frac{\partial U}{\partial \varphi},
\qquad
\Phi=\frac{\Phi_0}{2\pi}\,\varphi,
\label{eq:I_Phi_local}
\end{equation}
which, linearized around the minimum \(\varphi_{\min}\), gives
\begin{equation}
\delta I=
\frac{2\pi}{\Phi_0}\,
U''(\varphi_{\min})\,\delta\varphi,
\qquad
\delta\Phi=
\frac{\Phi_0}{2\pi}\,\delta\varphi.
\label{eq:linearized_IPhi}
\end{equation}
The corresponding local differential inductance of the biased \textit{rf}-SQUID cell is therefore
\begin{equation}
L_{\rm cell}
=
\frac{\delta\Phi}{\delta I}
=
\frac{(\Phi_0/2\pi)^2}{U''(\varphi_{\min})}
=
\frac{(\Phi_0/2\pi)^2}{E_{J_1}\,c_2}.
\label{eq:Lcell_c2}
\end{equation}
A locally stable operating point requires \(c_2>0\). The cubic coefficient \(c_3\) measures the asymmetry of the local well and is the coefficient responsible for 3WM, whereas \(c_4\) controls the leading quartic Kerr-type correction and therefore the residual 4WM contribution.
The design objective pursued in the following is precisely to identify operating points for which \(c_4\) vanishes while \(c_3\) remains appreciable and \(c_2\) stays positive.

Two limiting cases are worth noting. In the trivial limit \(\alpha\to 0\), one has \(\tau\to 0\), the Josephson contribution becomes phase independent, and the nonlinear response disappears, with \(c_2\to 1/\beta\) and \(c_{k\ge 3}\to 0\). 
Physically, this limit corresponds to consider the junction $1$ almost as a superconducting short, so that the effective element reduces to the junction $2$.
At the opposite end, when \(\alpha\to 1\), the effective element is maximally non-sinusoidal and the local cubic and quartic terms can become sizable, especially when the operating point approaches \(\varphi_{\min}\simeq \pi\). The maps of \(c_3\) and \(c_4\) obtained from Eqs.~\eqref{eq:c3} and \eqref{eq:c4} will therefore define the \emph{Kerr-free design ridge} of the elementary cell.

Figure~\ref{fig:c3c4-maps} shows how the cubic and quartic coefficients vary across the plane spanned by the asymmetry parameter \(\alpha\) and the reduced flux \(\phi\). While \(c_3\) displays a broad region of appreciable magnitude, \(c_4\) changes sign across a well-defined curve in parameter space. The condition \(c_4=0\) therefore selects a \emph{Kerr-free ridge}, marked with a black dot-dashed curve in panel~(b), along which the leading quartic contribution vanishes, while panel~(a) shows that the cubic term can remain finite and sizable on the same manifold. This separation between the cubic and quartic contributions is the basic mechanism exploited in the following to define the operating region of the cell: one should select bias points on the ridge \(c_4=0\) where \(|c_3|\) remains sufficiently large, while also retaining local stability through the condition \(c_2>0\). In this sense, the Kerr-free ridge is not merely a geometric feature of the parameter space, but the natural design manifold for the elementary \textit{rf}-SQUID cell.

\section{Kerr-free operating ridge}
\label{sec:kfr}

The local expansion developed above provides a direct design criterion for Kerr-free 3WM operation. For fixed \(\beta\), the condition
\begin{equation}
c_4(\alpha,\phi;\beta)=0
\label{eq:c4zero}
\end{equation}
defines, whenever a solution exists, a one-dimensional manifold in the \((\alpha,\phi)\) plane. 
We denote by
\begin{equation}
\phi^\star(\alpha;\beta)
\label{eq:phistar}
\end{equation}
the reduced flux bias that solves Eq.~\eqref{eq:c4zero} as a function of \(\alpha\), at a given \(\beta\).
This curve identifies the \emph{Kerr-free operating ridge} of the elementary cell, namely the set of bias points at which the leading quartic contribution to the local potential vanishes while the cubic term may remain finite. In this sense, the Kerr-free ridge may be viewed as a continuous generalization of the isolated sweet spots often discussed in the literature on single-element three-wave-mixing devices~\cite{RezaFrolov2026}.

To compare different bias points along this ridge, it is useful to introduce the cubic-to-stiffness ratio
\begin{equation}
\mathcal{F}(\alpha;\beta)=\frac{|c_3(\alpha,\phi^\star;\beta)|}{c_2(\alpha,\phi^\star;\beta)}.
\label{eq:FOM}
\end{equation}
This quantity should not be interpreted as a full amplifier performance metric, but rather as a local indicator of the balance between the useful cubic nonlinearity and the small-signal stiffness of the cell. Large values of \(\mathcal{F}\) correspond, at the level of the elementary-cell expansion, to operating points where the cubic response is enhanced relative to the local linear scale.

\begin{figure}[t]
  \centering
  \includegraphics[width=\columnwidth]{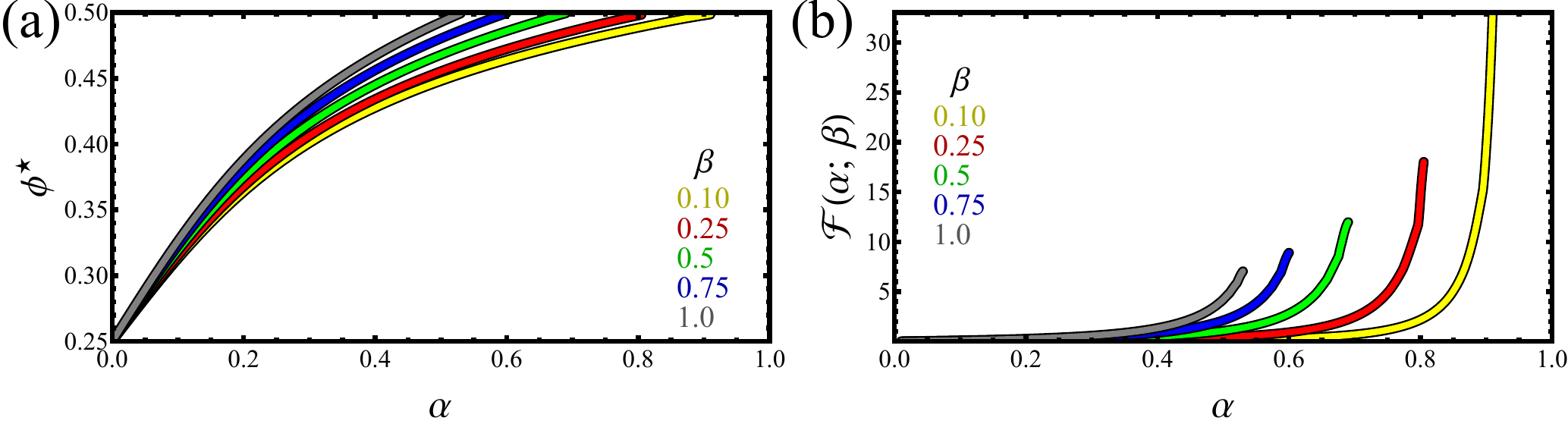}
\caption{Kerr-free operating ridge and cubic-to-stiffness ratio. Panel (a): \(\phi^\star(\alpha;\beta)\) solving \(c_4=0\) for \(\beta\in\{0.10,0.25,0.5,0.75,1\}\), i.e., the zero-anharmonicity ridge in the \((\alpha,\phi)\) plane. Panel (b): \(\mathcal{F}(\alpha;\beta)=|c_3|/c_2\) evaluated along the same ridge. In the parameter range explored here, the design trend is to move toward the largest feasible \(\alpha\), subject to \(c_2>0\), matching, phase-matching, and fabrication constraints.}
  \label{fig:c4ridge}
\end{figure}

The resulting Kerr-free ridge and the corresponding cubic-to-stiffness ratio are shown in Fig.~\ref{fig:c4ridge}. 
Panel~(a) reports \(\phi^\star(\alpha;\beta)\) for several values of \(\beta\), with \(\beta\) defined in Eq.~\eqref{eq:beta}. In the present normalized description, changing \(\beta\) is most naturally interpreted as varying the geometric inductance \(L\) of the superconducting loop at fixed \(I_{c_1}\), and therefore as tuning the relative weight of the inductive confinement with respect to the Josephson nonlinearity of the cell. 
Accordingly, increasing \(\beta\) softens the inductive contribution to the potential, so that the local landscape becomes more strongly shaped by the non-sinusoidal Josephson term; the Kerr-free condition \(c_4=0\) is then reached at higher flux bias. Consistently, panel~(a) shows that, for fixed \(\alpha\), the Kerr-free ridge shifts to larger \(\phi^\star\) as \(\beta\) increases.
At the same time, larger \(\beta\) makes the ridge approach \(\phi^\star \simeq 0.5\) already at smaller \(\alpha\), thereby reducing the accessible range of asymmetry along the Kerr-free manifold.

Panel~(b) shows \(\mathcal{F}(\alpha;\beta)\) evaluated along the same ridge. In the parameter range explored here, \(\mathcal{F}\) increases monotonically with \(\alpha\) for all the considered values of \(\beta\), indicating a clear trend toward larger asymmetry. For larger \(\beta\), the increase becomes steeper and shifts to smaller \(\alpha\), consistently with the fact that a softer inductive contribution enhances the relative weight of the non-sinusoidal Josephson term in the local expansion. Since \(\mathcal{F}\) combines both \(|c_3|\) and \(c_2\), this trend should be interpreted as a change in the local balance between cubic response and small-signal stiffness. At the same time, the termination of the curves shows that this trend cannot be pursued indefinitely, since the physically relevant operating region extends only as long as local stability is preserved and, as discussed below, impedance and implementation constraints remain satisfied.

The relevance of the present analysis for JTWPA design lies primarily at the level of the elementary nonlinear cell. For fixed SQUID parameter \(\beta\), the Kerr-free ridge identifies the physically relevant manifold of bias points along which the quartic contribution vanishes, while the local balance between stiffness and cubic response is controlled by the asymmetry parameter that controls the transparency  \(\alpha\) and the external magnetic flux \(\phi\). In this sense, transparency tuning shifts an essential part of the nonlinear design from loop asymmetry to the Josephson element itself. A full device-level treatment, including dispersion engineering, phase matching, gain performance and a full dynamical simulation of the two JJ model of Fig.~\ref{fig:device}(a), lies beyond the scope of the present paper.

\section{Passive matching and frequency-dependent constraints}
\label{sec:matching}

The present analysis has so far focused on the local nonlinear design of the elementary cell. Any prospective implementation, however, must also remain compatible with a realistic transmission-line environment. We therefore complement the Kerr-free cell analysis with a minimal passive-matching discussion, aimed at identifying which portion of the Kerr-free ridge remains physically accessible once linear circuit constraints are taken into account.

Following the lumped-element treatment of rf-SQUID-based JTWPAs in Ref.~\cite{OPeatain2023}, we model the elementary cell as a symmetric \(\pi\)-section. The series branch consists of the geometric loop inductance \(L_G\) in parallel with the differential inductance \(L_J\) of the transparent Josephson element, together with the system capacitance \(C_J\), while the shunt branch is a capacitance \(C_0\) to ground, see Fig.~\ref{fig:device}(b). The corresponding impedances are
\begin{equation}
Z_1(\omega)=
\left(
i\omega C_J+\frac{1}{i\omega L_{\rm eq}}
\right)^{-1}
\qquad\text{and}\qquad
Z_2=\frac{1}{i\omega C_0},
\label{eq:Z1Z2}
\end{equation}
with $L_{\rm eq}=\frac{L_G L_J}{L_G+L_J}$ being being the effective series inductance of the cell, see Fig.~\ref{fig:device}(c). Here \(L_J \equiv L_J(\alpha,\varphi_{\min})\) denotes the differential inductance of the transparent Josephson element, evaluated at the local operating point selected by the Kerr-free ridge.
Within the low-frequency small-signal reduced model adopted here, \(C_J\) denotes the effective lumped capacitance of the composite series element. This representation is expected to remain valid as long as the operating frequencies lie sufficiently below the internal plasma and relaxation scales of the two-junction subsystem, as described in Sect. \ref{sec:effective_element}. A full finite-frequency description would instead retain the full dynamics, including resistive and capacitive terms, of the two JJs, thus restoring the dynamics of the internal phase.

The differential inductance of the transparent Josephson element is obtained from the local slope of its CPR. Using Eq.~\eqref{eq:effJJ}, the effective CPR can be written as
\begin{equation}
I_{\blacktriangleright\!\blacktriangleleft}(\varphi)
=
I_{c_1}\,\frac{\alpha\sin\varphi}{\sqrt{1+\alpha^2+2\alpha\cos\varphi}}=
I_{c_1}\,\frac{\alpha\sin\varphi}{\sqrt{D(\varphi)}}.
\label{eq:Ieff_compact}
\end{equation}
The corresponding differential inductance is then defined by
\begin{equation}
\frac{1}{L_J(\alpha,\varphi)}
=
\frac{dI_{\blacktriangleright\!\blacktriangleleft}}{d\Phi}
=
\frac{2\pi}{\Phi_0}\,\frac{dI_{\blacktriangleright\!\blacktriangleleft}}{d\varphi}
=
\frac{2\pi I_{c_1}}{\Phi_0}\,
\left[
\frac{\alpha\cos\varphi}{\sqrt{D}}
+
\frac{\alpha^2\sin^2\varphi}{D^{3/2}}
\right].
\label{eq:LJ_def}
\end{equation}
Evaluating this expression at the operating point \(\varphi=\varphi_{\min}\) yields the small-signal differential inductance of the junction element:
\begin{equation}
\frac{1}{L_J(\alpha,\varphi_{\min})}
=
\frac{2\pi I_{c_1}}{\Phi_0}
\left[
\frac{\alpha\cos\varphi_{\min}}{\sqrt{D_{\min}}}
+
\frac{\alpha^2\sin^2\varphi_{\min}}{D_{\min}^{3/2}}
\right],
\quad\text{with}\quad
D_{\min}=1+\alpha^2+2\alpha\cos\varphi_{\min}.
\label{eq:LJtransparent}
\end{equation}
Thus, once \(\varphi_{\min}\) is fixed by the Kerr-free condition, the linear inductive part of the cell is completely determined by \(\alpha\), \(\beta\), and the geometric inductance \(L_G\).

Following the \(\pi\)-cell analysis of Ref.~\cite{OPeatain2023}, the \emph{low-frequency limit} of the present cell takes the usual quasi-TEM form with image impedance
\begin{equation}
Z_\pi \simeq \sqrt{\frac{L_{\rm eq}}{C_0}}.
\label{eq:Zpi_low}
\end{equation}

Imposing passive matching to a target line impedance \(Z_0=50~\Omega\) fixes the value that the ground capacitance \(C_0\) must take. Denoting this matched value by \(C_0^{\rm match}\), one obtains
\begin{equation}
C_0^{\rm match}(\alpha)
=
\frac{L_{\rm eq}(\alpha,\varphi_{\min})}{Z_0^2}
=
\frac{1}{Z_0^2}\,
\frac{L\,L_J(\alpha,\varphi_{\min})}{L+L_J(\alpha,\varphi_{\min})}.
\label{eq:C0rule}
\end{equation}
This relation already shows that the Kerr-free design trend toward larger \(\alpha\) cannot be pursued without bound. 
Indeed, the effective cell inductance in Eq.~\eqref{eq:Z1Z2} diverges when
\begin{equation}
L_G+L_J\left(\alpha,\varphi_{\min}\right)=0,
\label{eq:alphacrit}
\end{equation}
which defines a \emph{critical asymmetry parameter}, \(\alpha_{\rm crit}\), along the Kerr-free ridge. At this point the required ground capacitance also diverges, while for \(\alpha>\alpha_{\rm crit}\) the linearized cell would require a negative \(C_0^{\rm match}\), and therefore no passive \(50~\Omega\) implementation is possible.

\begin{figure}[t]
  \centering
  \includegraphics[width=\columnwidth]{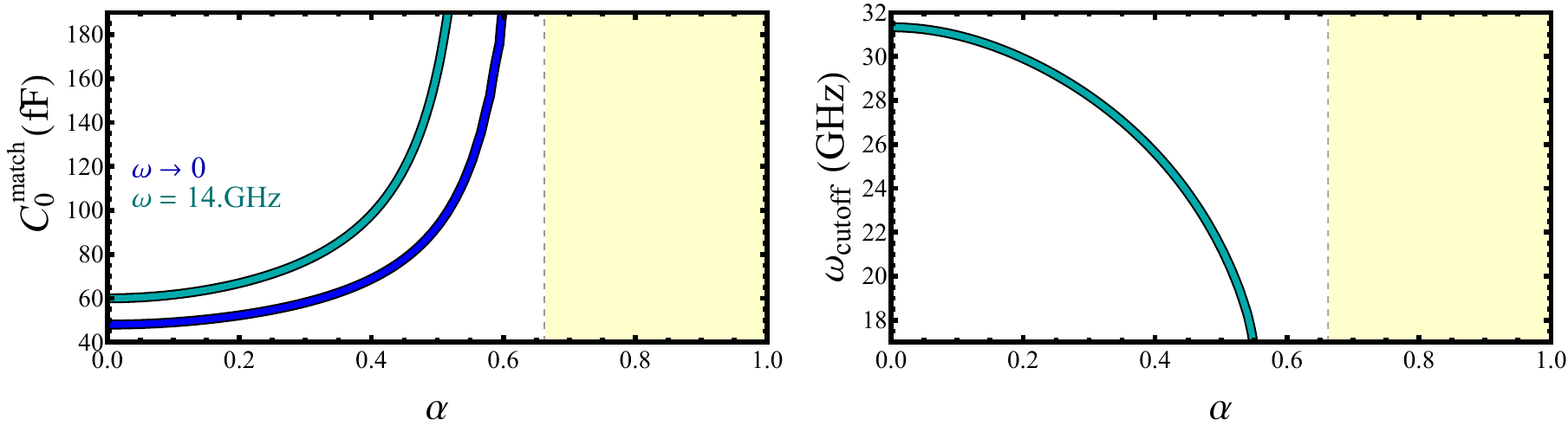}
\caption{Frequency-dependent matching, at fixed \(\beta=0.73\), along the Kerr-free ridge. 
Panel (a): ground capacitance \(C_0^{\rm match}(\alpha)\) obtained from both Eqs.~\eqref{eq:C0rule} and \eqref{eq:C0_omega}, at a given $\omega=14\;\text{GHz}$. 
Panel (b): cutoff frequency \(\omega_{\rm cutoff}(\alpha)\) from Eq.~\eqref{eq:cutoff_exact}. The gray dashed vertical line marks \(\alpha_{\rm crit}\); the shaded region corresponds to \(\alpha>\alpha_{\rm crit}\), where passive matching is no longer possible.}
\label{fig:Fig4_matching}
\end{figure}

Using Eq.~\eqref{eq:LJtransparent} together with the definition of the screening parameter in Eq.~\eqref{eq:beta}, the critical condition in Eq.~\eqref{eq:alphacrit} can be rewritten in the dimensionless form
\begin{equation}
\frac{\alpha\cos\varphi_{\min}}{\sqrt{D_{\min}}}
+
\frac{\alpha^2\sin^2\varphi_{\min}}{D_{\min}^{3/2}}
=
-\frac{1}{\beta}.
\label{eq:alphacrit_dimless}
\end{equation}
For a given \(\beta\), Eq.~\eqref{eq:alphacrit_dimless} alone does not determine the critical point, since both \(\alpha\) and \(\varphi_{\min}\) remain unknown. The Kerr-free condition \(c_4=0\) provides the second relation needed to identify the corresponding point on the Kerr-free manifold. The critical asymmetry \(\alpha_{\rm crit}\) is therefore obtained from the intersection of these two conditions, namely by solving Eq.~\eqref{eq:alphacrit_dimless} together with \(c_4=0\).
For the representative case \(\beta=0.73\), this gives
\[
\alpha_{\rm crit}\approx 0.66,
\qquad
\varphi_{\min}\approx 2.93.
\]

To include the role of the junction capacitance explicitly, we retain the full frequency dependence of the series branch. Equation~\eqref{eq:Z1Z2} can then be rewritten as
\begin{equation}
Z_1(\omega)
=
\frac{i\omega L_{\rm eq}}{1-\omega^2 L_{\rm eq}C_J},
\label{eq:Zseries_exact}
\end{equation}
from which the exact image impedance of the symmetric \(\pi\)-cell can be otained as
\begin{equation}
Z_\pi(\omega)=
\frac{\sqrt{Z_1\,Z_2}}{\sqrt{1+\tfrac{Z_1}{4Z_2}}}
=
\sqrt{\frac{L_{\rm eq}}{C_0\left[1-\omega^2 L_{\rm eq}\left(C_J+C_0/4\right)\right]}}.
\label{eq:Zpi_exact}
\end{equation}
In the limit \(\omega\to 0\), Eq.~\eqref{eq:Zpi_exact} reduces to Eq.~\eqref{eq:Zpi_low}, as expected. If instead the matching condition is imposed at a given frequency \(\omega\), one obtains
\begin{equation}
C_0^{\rm match}(\alpha,\omega)=
\frac{2}{\omega^2 L_{\rm eq}}
\left[
1-\omega^2 L_{\rm eq}C_J
-\sqrt{\bigl(1-\omega^2 L_{\rm eq}C_J\bigr)^2
-\frac{\omega^2 L_{\rm eq}^2}{Z_0^2}}
\right].
\label{eq:C0_omega}
\end{equation}
Since Eq.~\eqref{eq:C0_omega} follows from a quadratic condition, two mathematical roots are in principle possible. The physically acceptable one is the branch that gives \(C_0^{\rm match}>0\) and continuously recovers the DC result in Eq.~\eqref{eq:C0rule} as \(\omega\to 0\).

The \emph{lowest stop-band edge} is obtained by requiring the denominator of Eq.~\eqref{eq:Zpi_exact} to vanish, i.e.,
\begin{equation}
1-\omega^2 L_{\rm eq}\left(C_J+\frac{C_0}{4}\right)=0.
\label{eq:cutoff_condition}
\end{equation}
This gives the \emph{cutoff frequency of the \(\pi\)-cell},
\begin{equation}
\omega_{\rm cutoff}=
\frac{2}{\sqrt{L_{\rm eq}\left(C_0+4C_J\right)}}.
\label{eq:cutoff_exact}
\end{equation}
When the capacitance is fixed by matching at a given frequency, for example at the pump frequency \(\omega_p\), the cutoff in Eq.~\eqref{eq:cutoff_exact} is computed using the corresponding matched value \(C_0^{\rm match}(\alpha,\omega)\).

These constraints are illustrated in Fig.~\ref{fig:Fig4_matching} for a representative case at fixed \(\beta=0.73\). All quantities shown in Fig.~\ref{fig:Fig4_matching} are evaluated along the Kerr-free ridge, i.e., at \(\phi=\phi^\star(\alpha;\beta)\) and at the corresponding local minimum \(\varphi_{\min}(\beta,\alpha,\phi^\star)\). For each value of \(\alpha\), this determines \(L_J\) and hence \(L_{\rm eq}\). The matching capacitance \(C_0^{\rm match}\) is then obtained either from the DC condition in Eq.~\eqref{eq:C0rule} or, at finite frequency, from Eq.~\eqref{eq:C0_omega} at a given frequency, \(\omega\). In the illustrative example of Fig.~\ref{fig:Fig4_matching}, we use \(C_J=200~\mathrm{fF}\)~\cite{Pagano2022}.

As \(\alpha\) increases, the junction inductance decreases and approaches the singular condition \(L_J=-L_G\), which marks the critical asymmetry \(\alpha_{\rm crit}\). The approach to the critical condition is closely tied to the fact that, along the Kerr-free ridge, the corresponding flux bias \(\phi^\star\) moves very close to \(0.5\) as \(\alpha\) increases. Through the stationarity condition, this drives \(\varphi_{\min}\) toward \(\pi\), where the differential inductance of the transparent Josephson element becomes negative and eventually reaches the singular condition \(L_J=-L_G\). Panel~(a) reports the matching ground capacitance required for passive matching, comparing the DC condition of Eq.~\eqref{eq:C0rule} with the finite-frequency condition of Eq.~\eqref{eq:C0_omega} at \(14~\mathrm{GHz}\). In both cases, the required \(C_0^{\rm match}\) grows rapidly and diverges as \(\alpha\to\alpha_{\rm crit}\), while the finite-frequency correction remains quantitatively relevant, but does not modify the overall design trend. Panel~(b) shows the corresponding cutoff frequency \(\omega_{\rm cutoff}(\alpha)\), which decreases as \(\alpha\) increases because the effective inductance \(L_{\rm eq}\) of the cell grows strongly when \(L_J\) approaches \(-L_G\). Physically, this means that the line becomes increasingly inductive and the upper edge \(\omega_{\rm cutoff}\) of the first propagating band is pushed to lower frequencies, thereby reducing the useful spectral window available for amplifier operation. This means that, all relevant operating frequencies, including pump, signal, and idler tones, must remain sufficiently below \(\omega_{\rm cutoff}\) for the transmission-line description to remain physically meaningful.

The main implication is that the Kerr-free ridge identified from the local nonlinear expansion must be intersected with an independent linear realizability condition. Large \(\alpha\) is favorable from the point of view of the cubic response, but the accessible design space is ultimately truncated by passive matching and cutoff constraints before the formal Kerr-free trend can be pushed arbitrarily far.

\section{Conclusions}
\label{sec:conclusions}

We have presented a transparency-engineered route to Kerr-free three-wave mixing in rf-SQUID-based Josephson metamaterial cells.
The central result is that the non-sinusoidal Josephson energy (and hence current)-phase relation generated by an effective two-junction series element can be used as an explicit design resource for the local nonlinear response of the cell.

Within this framework, with a discussion of the underlying approximations, we have derived the flux- and asymmetry-dependent local expansion of the SQUID potential, identified the Kerr-free operating ridge defined by \(c_4=0\), and showed that this manifold supports operating points where the cubic response remains appreciable while local stability is preserved. In this sense, the proposed architecture makes the Kerr-free condition structurally explicit at the level of the elementary nonlinear cell.

A second central result is that the same mechanism that enhances the useful cubic response also brings the device closer to both linear implementation limits and nonlinear operating constraints. In particular, the accessible part of the Kerr-free ridge is ultimately truncated by the onset of passive-matching incompatibility, identified by the critical asymmetry \(\alpha_{\rm crit}\), together with the associated reduction of the cutoff frequency. The useful design space is therefore determined not only by the local nonlinear coefficients, but by the intersection of the Kerr-free manifold with stability and linear realizability constraints.

Taken together, these results establish the \emph{TRAIL}, i.e., the \emph{TRansparency-engineered and Asymmetry-tuned Inductive Loop}, as a compact building block for transparency-engineered three-wave-mixing Josephson metamaterials.
A full device-level treatment, including simulations based on the RCSJ model for each elementary junction, dispersion engineering, phase matching, and gain performance over a target band, lies beyond the scope of the present work and will be addressed in future investigations.





\nolinenumbers

\end{document}